\documentclass[amsmath,amssymb,10pt,eqsecnum, twocolumn]{revtex4}
\usepackage{graphicx, amsmath, amsmath, amssymb}
\usepackage[breaklinks, colorlinks, citecolor=blue]{hyperref}
\usepackage{natbib,hyperref,ifthen} 
\begin{document}
\title{Evidence for massive neutrinos from CMB and lensing observations}
\author{Richard A. Battye}
\email{richard.battye@manchester.ac.uk}
\affiliation{Jodrell Bank Centre for Astrophysics, School of Physics and Astronomy, University of Manchester, Manchester, M13 9PL, U.K.}
\author{Adam Moss}
\email{adam.moss@nottingham.ac.uk}
\affiliation{Centre for  Astronomy \& Particle Theory, University of Nottingham, University Park, Nottingham, NG7 2RD, U.K.}
\vskip 0.5cm
\begin{abstract}
We discuss whether massive neutrinos (either active or sterile) can reconcile some of the tensions within cosmological data that have been brought into focus by the recently released {\it Planck} data. We point out that a discrepancy is present when comparing the primary CMB and lensing measurements both from the CMB and galaxy lensing data using CFHTLenS, similar to that which arises when comparing CMB measurements and SZ cluster counts. A consistent picture  emerges and including a prior for the cluster constraints and BAOs we find that: for an active neutrino model with 3 degenerate neutrinos,  $\sum m_{\nu}= (0.320 \pm 0.081)\,{\rm eV}$, whereas for a sterile neutrino, in addition to 3 neutrinos with a standard hierarchy and $\sum m_{\nu}= 0.06\,{\rm eV}$,  $m_{\nu, \, \rm sterile}^{\rm eff}= (0.450 \pm 0.124)\,{\rm eV}$ and $\Delta N_{\rm eff} = 0.45 \pm 0.23$. In both cases there is a significant detection of modification to the neutrino sector from the standard model and in the case of the sterile neutrino it is possible to reconcile the BAO and local $H_0$ measurements. However, a caveat to our result is some internal tension between the CMB and lensing/cluster observations, and the  masses are in excess of those estimated from the shape of the matter power spectrum from galaxy surveys. 
\end{abstract}

\maketitle
Massive neutrinos are now part of the standard models of particle physics and cosmology. Solar and atmospheric neutrino experiments have measured two differences between the masses squared and from this it can be inferred that the sum of the active neutrino masses, $\sum m_{\nu}$, must be at least $0.06\,{\rm eV}$~\cite{2012JHEP...12..123G}. This is the quantity that can be constrained by cosmological observations. In addition, some experiments suggest that there could be a sterile neutrino that does not interact with the standard model~\cite{2013arXiv1306.6494C}, but in the context of cosmology still contributes a mass, $m_{\nu, \, \rm sterile}^{\rm eff}$, and in a model dependent way an increase in the number of effective relativistic degrees of freedom, $N_{\rm eff}=3.046+\Delta N_{\rm eff}$.

Using observations of the angular power spectrum of temperature anisotropies in the Cosmic Microwave Background (CMB) from the {\it Planck} satellite~\cite{2013arXiv1303.5062P}, polarisation measurements from the Wilkinson Microwave Anisotropy Probe (WMAP)~\cite{2012arXiv1212.5225B} and observations of Baryonic Acoustic Oscillations (BAOs)~\cite{Beut11,2012MNRAS.425..405B,2012MNRAS.427.2132P,2012MNRAS.427.3435A}, a constraint of $\sum m_{\nu}< 0.248\,{\rm eV}$ (95\% Confidence Level - CL) has been achieved in the case of active neutrinos~\cite{2013arXiv1303.5076P}, whereas in the sterile case $N_{\rm eff}<3.80$ and $m_{\nu, \, \rm sterile}^{\rm eff}<0.42\,{\rm eV}$ (95\% CL) for the case of a thermal sterile neutrino with  mass $<10\,{\rm eV}$. This analysis,  which will be referred to as {\it Planck} CMB+WP+BAO below,  was performed by adding $\sum m_{\nu}$ in the active case, or $m_{\nu, \, \rm sterile}^{\rm eff}$ and $N_{\rm eff}$ in the sterile case to the standard 6 parameter, ${\bf p}=\{\Omega_{\rm b}h^2, \Omega_{\rm c}h^2, \theta_{\rm MC}, A_{\rm S}, n_{\rm S}, \tau\}$, $\Lambda$CDM model. $\Omega_{\rm b}$ and $\Omega_{\rm c}$ are the baryonic and cold dark matter densities relative to the critical density. The Hubble constant is $100h\,{\rm km}\,{\rm sec}^{-1}\,{\rm Mpc}^{-1}$, which is a derived parameter; the parameter used in the fit is the acoustic scale, $\theta_{\rm MC}$. The primordial power spectrum is described by an amplitude, $A_{\rm S}$, and spectral index, $n_{\rm S}$. The optical depth to the epoch of reionization is $\tau$.
 
In addition, the results of {\it Planck} have highlighted a possible discrepancy between the cosmological parameters preferred  by CMB data and BAOs, and those which come from fitting the counts of galaxy clusters selected using the Sunyaev Zeldovich (SZ) effect~\cite{2013arXiv1303.5080P}. This is best quantified in terms of derived parameters $\Omega_{\rm m}=\Omega_{\rm b}+\Omega_{\rm c}$ and $\sigma_8$, which are the total matter density relative to critical and the amplitude of fluctuations on $8h^{-1}\,{\rm Mpc}$ scales, respectively, {and $\Omega_{\rm c}$ includes the neutrino contribution}. Using a bias between the hydrostatic mass and the true mass of $20\%$ ($1-b=0.8$ in the parlance of \cite{2013arXiv1303.5080P}) the SZ cluster counts require $\sigma_8(\Omega_{\rm m}/0.27)^{0.3}= 0.78 \pm 0.01$, which is lower than preferred by CMB data. A similar discrepancy can be inferred from other measurements of cluster number counts using the SZ~\cite{2013ApJ...763..147B,2013JCAP...07..008H}, X-rays~\cite{vik09} and optical richness~\cite{2010ApJ...708..645R}. This could be due to a number of incorrect assumptions in calculation of the cluster number counts which are in common between the different analyses, for example, the relationship between the observable and the true mass or mass function. However, it could also be as a result of additional physics that is missing from the standard 6 parameter model and in \cite{2013arXiv1303.5080P} it was suggested that massive active neutrinos could lead to an improved fit, with $\sum m_{\nu}=  (0.22 \pm 0.09)\,{\rm eV} $ from an analysis of CMB+SZ+BAO. Although not explicitly discussed there a similar effect could be achieved from the inclusion of sterile neutrinos.

\begin{figure} 
\centering
\mbox{\resizebox{0.48\textwidth}{!}{\includegraphics{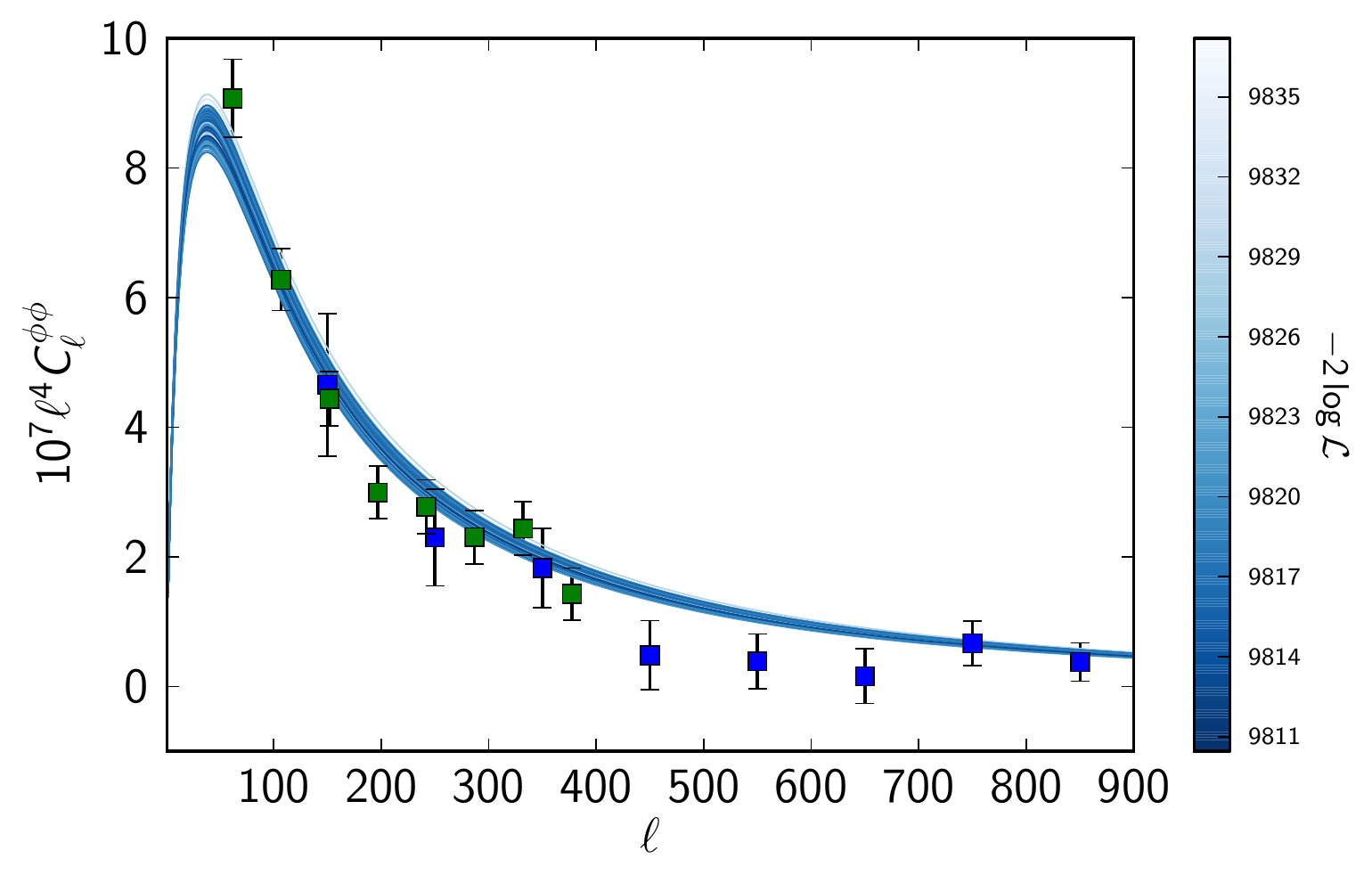}}}
\mbox{\resizebox{0.48\textwidth}{!}{\includegraphics{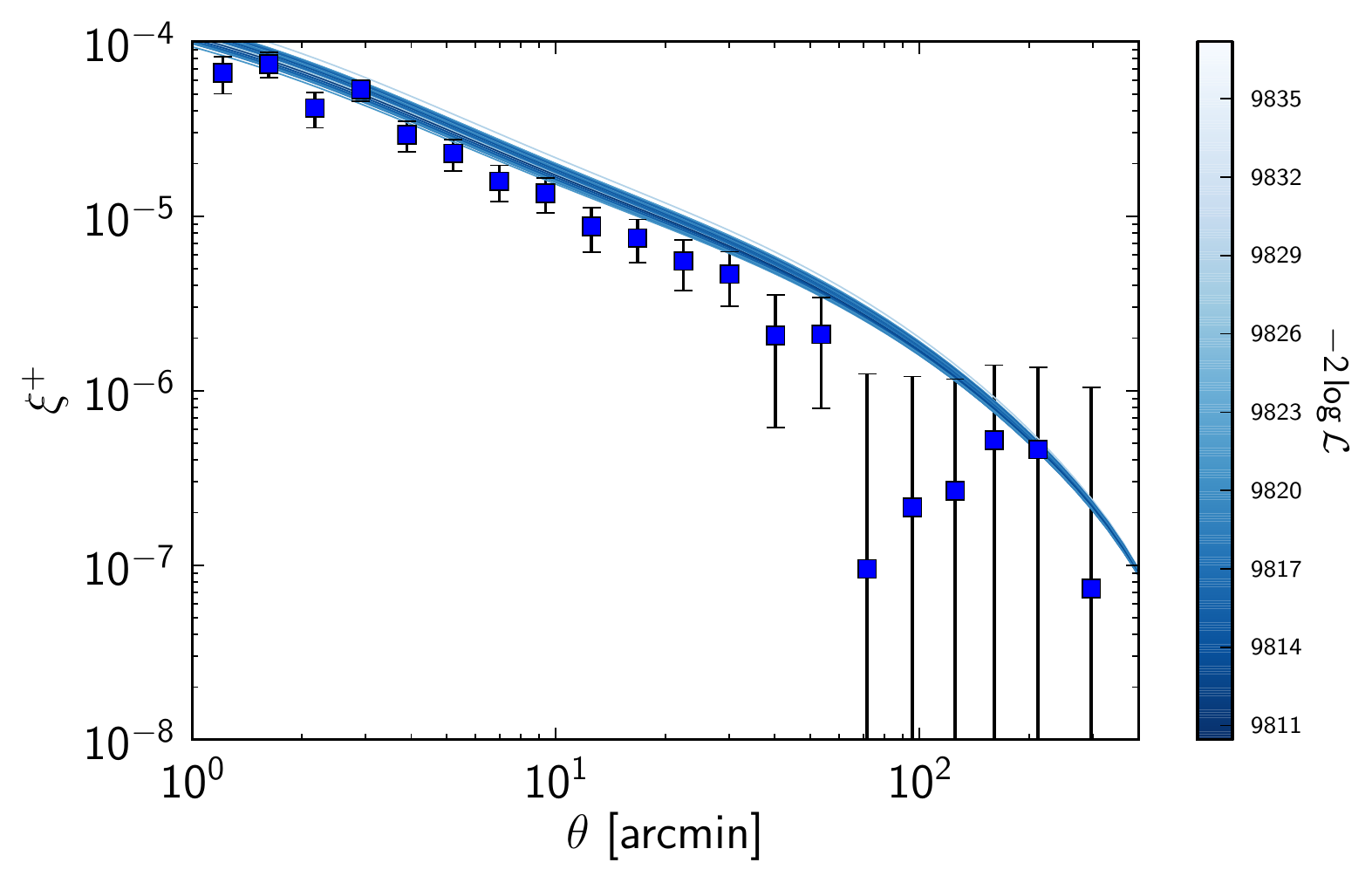}}}
\caption{The CMB lensing power spectrum (top) data points from {\it Planck} (green squares) and SPT (blue squares) and the shear correlation function $\xi^{+}$ from CFHTLenS (bottom), compared to predictions for parameters from samples of the {\it Planck} CMB+WP+BAO   MCMC chains ($\Lambda$CDM, zero neutrino mass) with non-linear corrections~\cite{Smith:2002dz,Takahashi:2012em}. In both cases, the data is systematically lower than theory, although the significance is somewhat lower than the eye would suggest in the case of CFHTLenS due to correlations between data points{, which ranges from $\sim10\%$ to $\sim50\%$ on small and large scales respectively.  SPT data has a similar level of correlation, and for  {\it Planck} the correlation is negligible.}}
\label{fig:spectra}
\end{figure}

In this {\it letter} we will make the case that this explanation of the discrepancy between the CMB and cluster counts is also favoured by lensing data. This data comes from CMB lensing as detected by {\it Planck}~\cite{2013arXiv1303.5077P} and the South Pole Telescope (SPT)~\cite{vanEngelen:2012va}, and also from galaxy lensing detected by the Canada-France-Hawaii Telescope Lensing Survey (CFHTLenS)~\cite{2013MNRAS.430.2200K}. A careful reading of ~\cite{2013arXiv1303.5077P}  and ~\cite{2013MNRAS.430.2200K} might already suggest this: the increase in the limit $\sum m_{\nu}$ for active neutrinos from CMB lensing and the constraint of $\sigma_8(\Omega_{\rm m}/0.27)^{0.59}= 0.79\pm 0.03 $ from the CFHTLenS are both symptoms of this tension. A simple illustration of this point is to just compare the expected lensing spectra for the best fitting models to {\it Planck} CMB+WP+BAO reported in \cite{2013arXiv1303.5076P}. In Fig.~\ref{fig:spectra} we have plotted the measurements of the CMB lensing power spectrum, $C_{\ell}^{\phi\phi}$, and the galaxy lensing correlation function, $\xi^{+}(\theta)$ (the Hankel transform of the convergence power spectrum $P_{\kappa}$), along with model predictions colour coded by their likelihood. It is clear that, in both cases, those parameter combinations that are a good fit to the CMB+BAO data predict a higher level of lensing correlations than observed ($\Delta\chi^2 \sim 20$), indicating that there could be something missing within the model. We will make this explicit by performing a full joint likelihood analysis of the publicly available lensing data and combining this with a prior on $\sigma_8(\Omega_{\rm m}/0.27)^{0.3}$ coming from the SZ cluster counts, which will lead to a significant preference for such models. We note that this is not equivalent to performing a full joint analysis including the SZ likelihood -- which is not publicly available -- but we have tested that this leads to similar results to those presented in \cite{2013arXiv1303.5080P}.

There are two separate analyses that we have performed: a model with the standard six parameters, ${\bf p}$, and 1 extra parameter $\sum m_{\nu}$ with $N_{\nu}=3$ ($N_{\nu}$ is the number of massive neutrinos) and $N_{\rm eff}=3.046$;  a model with a total of 8 parameters -- ${\bf p}+\{m_{\nu, \, \rm sterile}^{\rm eff},N_{\rm eff}\}$ -- and $\sum m_{\nu}=0.06\,{\rm eV}$, $N_{\nu}=1$ ({the other neutrinos in the standard hierarchy are massless}). The first represents a degenerate active neutrino scenario, that is appropriate for large values of $\sum m_{\nu}$, whereas the second is a sterile neutrino scenario with active neutrinos  in a standard hierarchy that has the lowest value of $\sum m_{\nu}$ allowed by the solar and atmospheric constraints on the mass differences.

 In both cases we will follow the procedure outlined in \cite{2013arXiv1303.5080P}  and use the {\it Planck} likelihood~\cite{2013arXiv1303.5075P} that includes a number of nuisance parameters describing the contamination from our own galaxy, extragalactic sources and the SZ effect. We will consider three data combinations: (I) {\it Planck} CMB+WP+BAO; (II) {\it Planck} CMB+WP+BAO+lensing where lensing is both the CMB lensing from {\it Planck } and SPT and galaxy lensing from CFHTLenS; (III) {\it Planck} CMB+WP+BAO+lensing+SZ cluster counts imposed using a prior in the $\sigma_8-\Omega_{\rm m}$ plane of $\sigma_8(\Omega_{\rm m}/0.27)^{0.3}= 0.78 \pm 0.01$. For CFHTLenS we use the $\xi^{\pm}$ correlation functions and covariance matrix as described in~\cite{2013MNRAS.430.2200K}, choosing the smallest and largest angular scales to be 0.9 and 300 arcmin respectively.  To compare the shear with that measured from large scale structure we correct the power spectrum on non-linear scales using the {\tt Halofit} fitting formulae~\cite{Smith:2002dz,Takahashi:2012em}. For SPT lensing data we follow the same procedure as in~\cite{vanEngelen:2012va}, rescaling the diagonals of the covariance matrix according to sample variance, and adding an additional calibration-induced uncertainty to the covariance. 
 
\begin{figure} 
\centering
\mbox{\resizebox{0.4\textwidth}{!}{\includegraphics{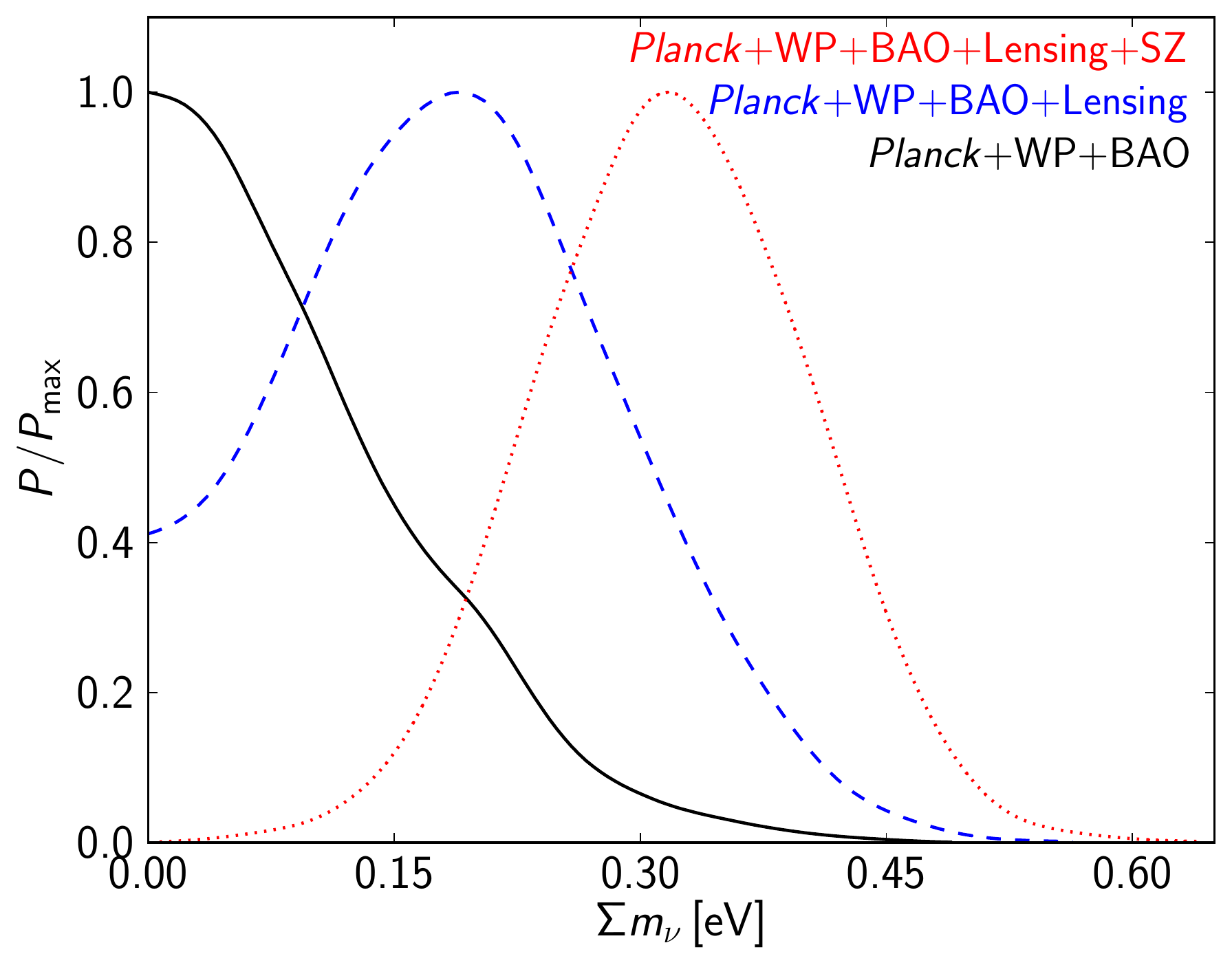}}}
\caption{Marginalized likelihoods for $\sum m_{\nu}$.  The datasets are colour coded in the legend, but the solid line is for (I), the dashed line is for (II) and the dotted line is for (III). It is clear that inclusion of lensing leads to a preference for $\sum m_{\nu}>0$ which is compatible with that coming from the SZ cluster counts and that there is a strong preference $(\approx 4\sigma)$ in the case of dataset (III).}
\label{fig:active-1D}
\end{figure}

\begin{figure} 
\centering
\mbox{\resizebox{0.4\textwidth}{!}{\includegraphics{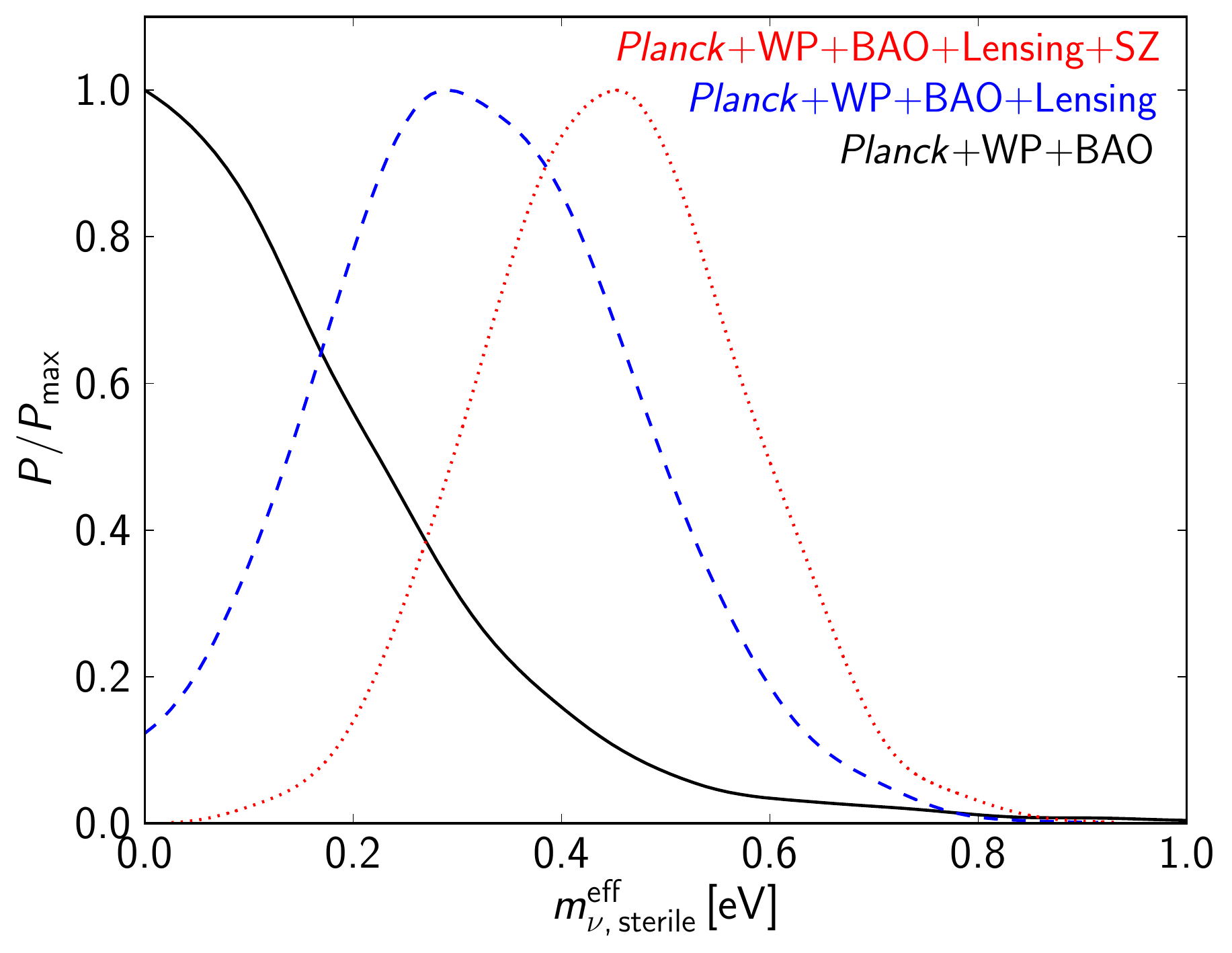}}}
\mbox{\resizebox{0.4\textwidth}{!}{\includegraphics{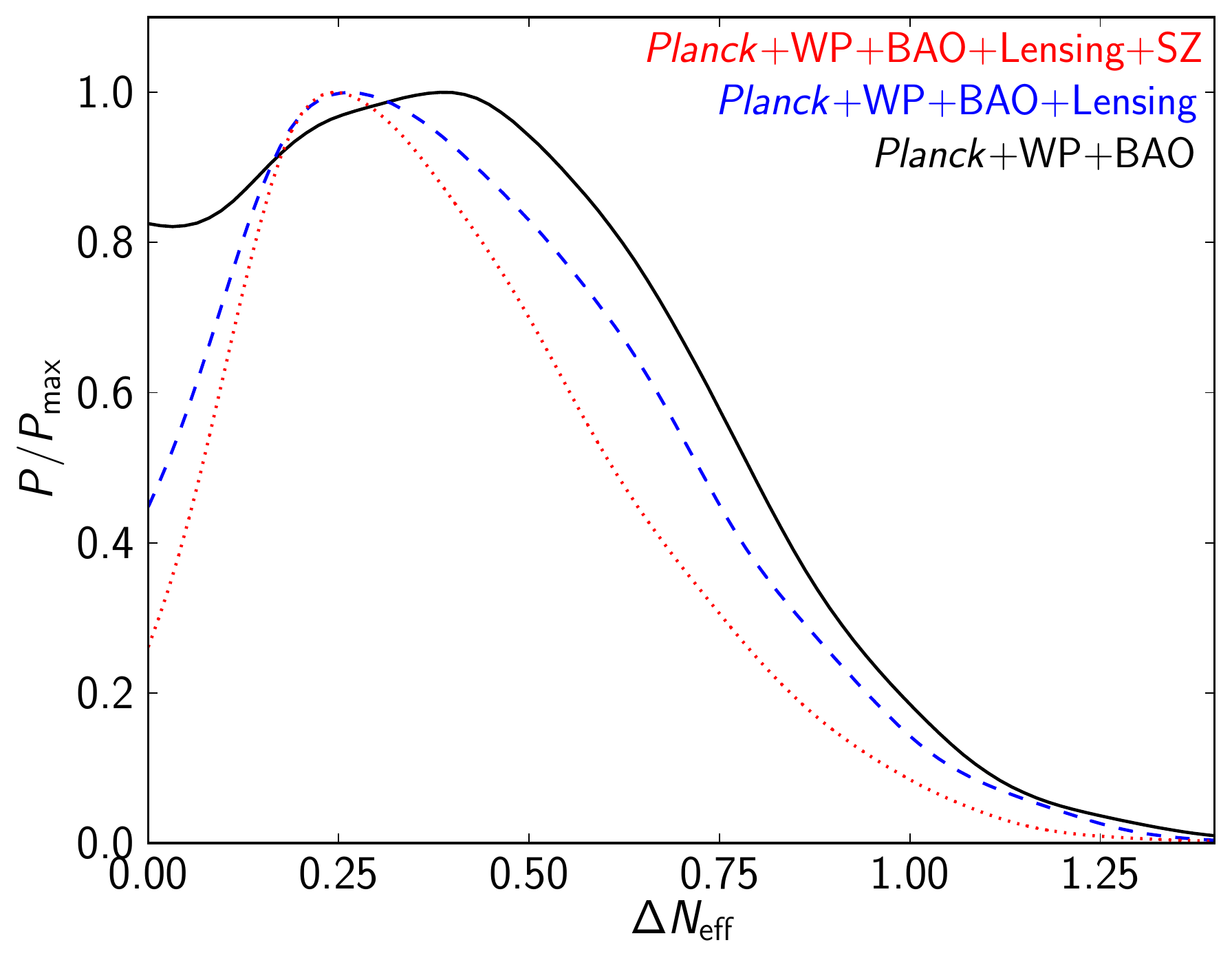}}}
\mbox{\resizebox{0.4\textwidth}{!}{\includegraphics{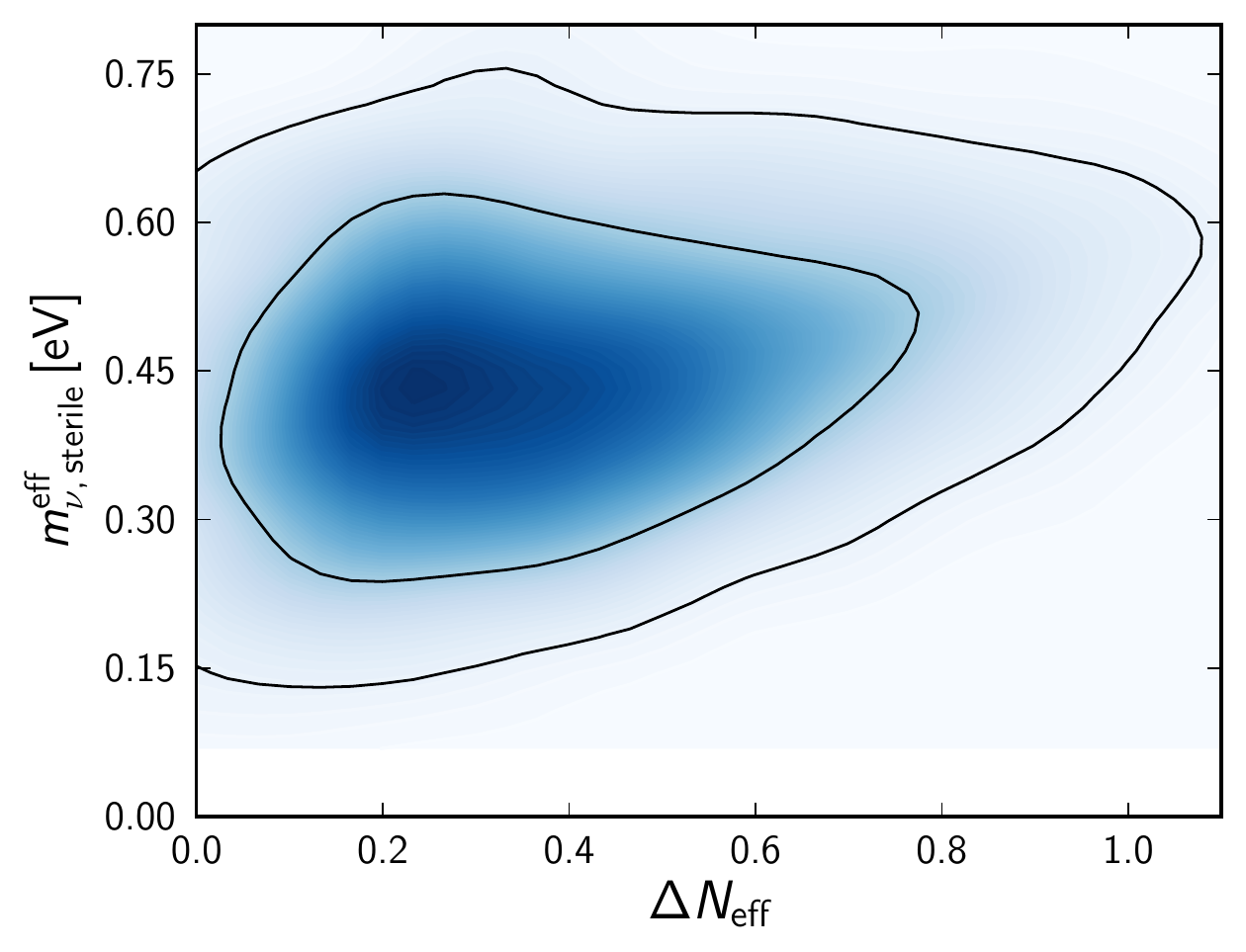}}}
\caption{Marginalized likelihoods for the sterile neutrino mass and the extra effective degrees of freedom (top and middle panels, labelling as in Fig.~\ref{fig:active-1D}), together with the 2D joint likelihood (bottom panel).}
\label{fig:ste_plots}
\end{figure}

\begin{table*}[!htb]
\label{tab:all}
\begin{center}
\begin{tabular}{|c|c|c|c|c|c|c|} \hline

  &  \multicolumn{3}{|c|}{Active neutrinos} &  \multicolumn{3}{|c|}{Sterile neutrinos} \\ \hline 

 Parameter & I & II & III &  I & II & III  \\ \hline

$\Omega_{\rm b} h^2 $ & $0.02218 \pm 0.00025$ & $0.02231 \pm 0.00024$ &$0.02234 \pm 0.00024$  & $0.02244 \pm 0.00029$ & $0.02256 \pm 0.00028$ &  $0.02258 \pm 0.00027$   \\ \hline
$\Omega_{\rm c} h^2 $ & $0.1184 \pm 0.0018$ & $0.1162 \pm 0.0013$  & $0.1152 \pm 0.0013$ & $0.1244 \pm 0.0051$ & $0.1221 \pm 0.0041$ & $0.1206 \pm 0.0040$ \\ \hline
$100\theta _{\rm MC}$ & $1.04151 \pm 0.00056$ & $1.04163 \pm 0.00056$ & $1.04170 \pm 0.00056$  & $1.04086 \pm 0.00072$ & $1.04106 \pm 0.00065$ & $1.04117 \pm 0.00065$  \\ \hline
$\tau_{\rm R} $ & $0.092 \pm 0.013$ & $0.093 \pm 0.013$ & $0.096 \pm 0.014$ & $0.096 \pm 0.014$ & $0.099 \pm 0.014$ & $0.097 \pm 0.014$   \\ \hline
$n_{\rm S} $ & $0.9643 \pm 0.0059$ & $0.9685 \pm 0.0052$ & $0.9701 \pm 0.0056$  & $0.9775 \pm 0.0106$ & $0.9792 \pm 0.0106$ &  $0.9772 \pm 0.0104$  \\ \hline
$\log (10^{10} A_{\rm S})$ & $3.091 \pm 0.025$ & $3.088 \pm 0.024$ &  $3.091 \pm 0.026$ & $3.115 \pm 0.030$ & $3.116 \pm 0.031$ & $3.109 \pm 0.030$  \\ \hline \hline
$\sum m_{\nu}$ [eV] & $< 0.254$ & $<0.358$ & $0.320 \pm 0.081$ & - & - & - \\ \hline 
$m_{\nu, \, \rm sterile}^{\rm eff}$ [eV] & - & -& -& $<0.479$ & $0.326 \pm 0.143$ & $0.450 \pm 0.124$ \\ \hline 
$\Delta N_{\rm eff} $ & - & - & - & $<0.98$ & $<0.96$ &  $0.45 \pm 0.23$ \\ \hline \hline
$H_0  $ & $67.65 \pm 0.90$ & $67.80 \pm 1.08$ & $67.00 \pm 1.07$ & $69.69 \pm 1.68$ & $69.51 \pm 1.41$ &   $69.02 \pm 1.21$  \\ \hline 
$\Omega_{\rm m}$ & $0.310 \pm 0.12$ & $0.306 \pm 0.13$ & $0.314 \pm 0.13$ & $0.308 \pm 0.12$ & $0.308 \pm 0.12$ & $0.312 \pm 0.12$ \\ \hline
$\sigma_{8}$ & $0.818 \pm 0.023$ & $0.789 \pm 0.020$ & $0.757 \pm 0.014$ & $0.813 \pm 0.032$ & $0.779 \pm 0.020$ & $0.756 \pm 0.012$ \\ \hline \hline
$- 2 \ln \mathcal{L}_{\rm CMB}  $ & 9804.96 & 9808.41 & 9811.35 & 9804.69 & 9809.15 & 9809.09 \\ \hline 
$- 2 \ln \mathcal{L}_{\rm BAO}  $ & 1.38  & 3.09 & 1.29  & 1.62 & 1.61 & 1.99 \\ \hline 
$- 2 \ln \mathcal{L}_{\rm Lensing}  $ & $-1009.56^{\star}$ & -1030.12 & -1030.05 & $-1018.68^{\star}$ &-1031.76  & -1031.43 \\ \hline 
$- 2 \ln \mathcal{L}_{\rm SZ}  $ & $92.49^{\star}$ & $5.61^{\star}$ & 2.19 & $59.62^{\star}$ & $5.74^{\star}$ & 0.37 \\ \hline 
$- 2 \ln \mathcal{L} $ & 9806.34 & 8781.37 & 8784.78 & 9806.31 & 8779.00 & 8780.02\\ \hline 

\end{tabular}
\end{center}
\caption{Summary of parameter constraints for both the active and sterile neutrino analyses discussed in the text. Likelihoods denoted by $^{\star}$ are not included in the total likelihood for that particular dataset.}
\label{tab:main}
\end{table*}

 Detailed constraints on the parameters are presented in table~\ref{tab:main}. We first turn our attention to the active neutrino case. In Fig.~\ref{fig:active-1D} we present the 1D likelihood for $\sum m_{\nu}$ which illustrates that the upper bound of $\sum m_{\nu}<0.254\,{\rm eV}$ that we find in the case of (I) is weakened by the inclusion of the lensing data and that a peak develops in the likelihood at non-zero $\sum m_{\nu}$. By itself the lensing data is not sufficiently strong to induce a strong preference, but the inclusion of the prior from the SZ cluster catalogue leads to $\sum m_\nu = (0.320\pm 0.081)\,{\rm eV}$, which corresponds to $\approx 4\sigma$ detection of $\sum m_{\nu}>0$.

We now consider the sterile neutrino model which leads to a similar, but even stronger result. The results are present in Fig.~\ref{fig:ste_plots}. For (II) we find that there is a $2.3\sigma$ preference for $m_{\nu, \, \rm sterile}^{\rm eff}>0$ with $m_{\nu, \, \rm sterile}^{\rm eff}=(0.326\pm 0.143)\,{\rm eV}$ although there is only an upper bound of $\Delta N_{\rm eff}<0.96$. This is strengthened to $m_{\nu, \, \rm sterile}^{\rm eff}=(0.450\pm 0.124)\,{\rm eV}$ and $\Delta N_{\rm eff}=0.45\pm 0.23$ for (III). 

The sterile neutrino model has the added feature that it can be made compatible with the direct measurement of Hubble's constant from Cepheid variables in nearby galaxies which appears to be at odds with the values of  inferred by CMB analyses~\cite{2013arXiv1307.7715W}. This is illustrated by preferred values of $H_0$ in these models presented in table~\ref{tab:main} being significantly larger than in the active neutrino model without leading to an increased $-2 \ln \mathcal{L}_{\rm BAO}$.  Including the prior $h = 0.738 \pm 0.024$ from \cite{2011ApJ...730..119R} to (III) modifies the constraints to $\sum m_{\nu} = (0.246 \pm 0.077)\,{\rm  eV}$ in the active neutrino model and  $m_{\nu, \, \rm sterile}^{\rm eff} = (0.425 \pm 0.122)\,{\rm eV}$, $\Delta N_{\rm eff} = 0.592 \pm 0.275 $ in the sterile neutrino model, with $\Delta \chi^2 = 6.3 $ between the two. {The larger change in the mean value in the active case arises from the degeneracy between $H_0, \Omega_{\rm m}$ and $\sum m_{\nu}$, with increased $H_0$ corresponding to lower $\Omega_{\rm m}$ and $\sum m_{\nu}$.}

{It is also instructive to perform the same analysis with WMAP 9-year plus high-$\ell$ data in the place of {\it Planck}  for (III) (see~\cite{2013arXiv1303.5076P} for details of the high-$\ell$ analysis). We find $\sum m_{\nu} = (0.297 \pm 0.084)\,{\rm  eV}$ in the active neutrino model and  $m_{\nu, \, \rm sterile}^{\rm eff} = (0.367 \pm 0.156)\,{\rm eV}$, $\Delta N_{\rm eff} = 0.276 \pm 0.203$ in the sterile neutrino model. This increases our confidence that  {\it Planck} results are consistent with WMAP, but at higher significance.}

The main argument that we have presented in this paper is that  amplitude of Large-Scale Structure (LSS) when normalized to the amplitude of CMB fluctuations are in excess of that inferred by lensing and cluster counts, and indeed that these two measures of the amplitude of the power spectrum  are consistent. If we add massive neutrinos -- either active or sterile -- to the cosmological model then we get significant detections that are due to the decrease in small- relative to large-scale power in such models. These measures of the amplitude of LSS are not without their modelling difficulties, but the fact that they appear to agree is encouraging. There are, however, caveats to what we have said.

Firstly, we note that the improved global fit when including massive neutrinos is usually at the expense of an increase in $-2 \ln \mathcal{L}_{\rm CMB}$. This increase is $\approx 6.3$ for the best-fitting model in the active neutrino case and $\approx4.4$ for sterile neutrinos. This is outweighed by the significant reductions in $-2 \ln \mathcal{L}_{\rm Lensing}$ and $- 2 \ln \mathcal{L}_{\rm SZ}$ (see table~\ref{tab:main}), but is reflected by the fact that preferred values in the case of detections overlap somewhat the $95\%$ CL limits in the case of (I). {We have quantified this tension by performing a separate analysis for the active neutrino case. A Bayesian approach is to assume there 
are two neutrino masses in the MCMC, one for the CMB + BAO part of the likelihood, $\sum m_{\nu}^{\rm CMB+BAO}$,  and one for the LSS component, $\sum m_{\nu}^{\rm LSS}$, and they otherwise share the same cosmological parameters. We find the marginalised posterior in the case of (III) is $\sum m_{\nu}^{\rm LSS}-\sum m_{\nu}^{\rm CMB+BAO}>0$ at $2.8\sigma$.} It could be that there exists a variant of the massive  neutrino model that leads to a better fit to the CMB data while preserving the positive impact on the amplitude of LSS.

We also note that there are published limits on the $\sum m_{\nu}$ that are contrary to the arguments presented here~\cite{2010PhRvL.105c1301T,
2013arXiv1306.4153R}. These are based on the shape of the power spectrum of LSS as opposed to its amplitude. We believe that these constraints could easily be ignored if there were significant scale dependent bias in the galaxy populations detected by the redshift surveys. For example, it has been show~\cite{2008MNRAS.385..830S} that differing amounts of red and blue galaxies in surveys can make it difficult to use the shape to determine cosmological parameters. While we acknowledge the existence of these limits, our opinion is that they are much less reliable than the arguments that we have put forward.

{\it Note added for arXiv:} ~ As we were preparing to submit this paper, a preprint appeared on the arXiv~\cite{2013arXiv1308.3255H} closely related to this work. Despite the different treatment of CFHTLens data, the results are in excellent agreement. We had only just submitted our paper to the {\it Planck} editorial board for review, which took 8 days, explaining the delay in this work appearing.  

{\it Acknowledgements:} ~This research was supported by STFC. We thank Jim Zibin for useful discussions.  We acknowledge the use of the {\tt CAMB}~\cite{2000ApJ...538..473L} and {\tt COSMOMC}~\cite{2002PhRvD..66j3511L} codes, and  the use of {\it Planck} data. The development of {\it Planck} has been supported by: ESA; CNES and 
CNRS/INSU-IN2P3-INP (France); ASI, CNR, and INAF (Italy); NASA and DoE 
(USA); STFC and UKSA (UK); CSIC, MICINN, JA and RES (Spain); Tekes, AoF 
and CSC (Finland); DLR and MPG (Germany); CSA (Canada); DTU Space 
(Denmark); SER/SSO (Switzerland); RCN (Norway); SFI (Ireland); FCT/MCTES 
(Portugal); and PRACE (EU). A description of the {\it Planck} Collaboration 
and a list of its members, including the technical or scientific 
activities in which they have been involved, can be found at 
\url{http://www.sciops.esa.int/index.php?project=planck&page=Planck_Collaboration}

\bibliography{lensing}

\def\eprinttmppp@#1arXiv:@{#1}
\providecommand{\arxivlink[1]}{\href{http://arxiv.org/abs/#1}{arXiv:#1}}
\def\eprinttmp@#1arXiv:#2 [#3]#4@{\ifthenelse{\equal{#3}{x}}{\ifthenelse{
\equal{#1}{}}{\arxivlink{\eprinttmppp@#2@}}{\arxivlink{#1}}}{\arxivlink{#2}
  [#3]}}
\providecommand{\eprintlink}[1]{\eprinttmp@#1arXiv: [x]@}
\renewcommand{\eprint}[1]{\eprintlink{#1}}
\providecommand{\adsurl}[1]{\href{#1}{ADS}}
\renewcommand{\bibinfo}[2]{\ifthenelse{\equal{#1}{isbn}}{\href{http://cosmologist.info/ISBN/#2}{#2}}{#2}}
\begin{thebibliography}{28}
\expandafter\ifx\csname natexlab\endcsname\relax\def\natexlab#1{#1}\fi
\expandafter\ifx\csname bibnamefont\endcsname\relax
  \def\bibnamefont#1{#1}\fi
\expandafter\ifx\csname bibfnamefont\endcsname\relax
  \def\bibfnamefont#1{#1}\fi
\expandafter\ifx\csname citenamefont\endcsname\relax
  \def\citenamefont#1{#1}\fi
\expandafter\ifx\csname url\endcsname\relax
  \def\url#1{\texttt{#1}}\fi
\expandafter\ifx\csname urlprefix\endcsname\relax\def\urlprefix{URL }\fi

\bibitem[{\citenamefont{{Gonzalez-Garcia}
  et~al.}(2012)\citenamefont{{Gonzalez-Garcia}, {Maltoni}, {Salvado}, and
  {Schwetz}}}]{2012JHEP...12..123G}
\bibinfo{author}{\bibfnamefont{M.~C.} \bibnamefont{{Gonzalez-Garcia}}},
  \bibinfo{author}{\bibfnamefont{M.}~\bibnamefont{{Maltoni}}},
  \bibinfo{author}{\bibfnamefont{J.}~\bibnamefont{{Salvado}}},
  \bibnamefont{and}
  \bibinfo{author}{\bibfnamefont{T.}~\bibnamefont{{Schwetz}}},
  \bibinfo{journal}{Journal of High Energy Physics}
  \textbf{\bibinfo{volume}{12}}, \bibinfo{pages}{123} (\bibinfo{year}{2012}),
  \eprint{1209.3023}.

\bibitem[{\citenamefont{{Conrad} et~al.}(2013)\citenamefont{{Conrad}, {Louis},
  and {Shaevitz}}}]{2013arXiv1306.6494C}
\bibinfo{author}{\bibfnamefont{J.~M.} \bibnamefont{{Conrad}}},
  \bibinfo{author}{\bibfnamefont{W.~C.} \bibnamefont{{Louis}}},
  \bibnamefont{and} \bibinfo{author}{\bibfnamefont{M.~H.}
  \bibnamefont{{Shaevitz}}}, \bibinfo{journal}{ArXiv e-prints}
  (\bibinfo{year}{2013}), \eprint{1306.6494}.

\bibitem[{\citenamefont{{Planck Collaboration}
  et~al.}(2013{\natexlab{a}})\citenamefont{{Planck Collaboration}, {Ade},
  {Aghanim}, {Armitage-Caplan}, {Arnaud}, {Ashdown}, {Atrio-Barandela},
  {Aumont}, {Baccigalupi}, {Banday} et~al.}}]{2013arXiv1303.5062P}
\bibinfo{author}{\bibnamefont{{Planck Collaboration}}},
  \bibinfo{author}{\bibfnamefont{P.~A.~R.} \bibnamefont{{Ade}}},
  \bibinfo{author}{\bibfnamefont{N.}~\bibnamefont{{Aghanim}}},
  \bibinfo{author}{\bibfnamefont{C.}~\bibnamefont{{Armitage-Caplan}}},
  \bibnamefont{et~al.}, \bibinfo{journal}{ArXiv e-prints}
  (\bibinfo{year}{2013}{\natexlab{a}}), \eprint{1303.5062}.

\bibitem[{\citenamefont{{Bennett} et~al.}(2012)\citenamefont{{Bennett},
  {Larson}, {Weiland}, {Jarosik}, {Hinshaw}, {Odegard}, {Smith}, {Hill},
  {Gold}, {Halpern} et~al.}}]{2012arXiv1212.5225B}
\bibinfo{author}{\bibfnamefont{C.~L.} \bibnamefont{{Bennett}}},
  \bibinfo{author}{\bibfnamefont{D.}~\bibnamefont{{Larson}}},
  \bibinfo{author}{\bibfnamefont{J.~L.} \bibnamefont{{Weiland}}},
  \bibinfo{author}{\bibfnamefont{N.}~\bibnamefont{{Jarosik}}},
  \bibnamefont{et~al.}, \bibinfo{journal}{ArXiv e-prints}
  (\bibinfo{year}{2012}), \eprint{1212.5225}.

\bibitem[{\citenamefont{{Beutler} et~al.}(2011)\citenamefont{{Beutler},
  {Blake}, {Colless}, {Jones}, {Staveley-Smith}, {Campbell}, {Parker},
  {Saunders}, and {Watson}}}]{Beut11}
\bibinfo{author}{\bibfnamefont{F.}~\bibnamefont{{Beutler}}},
  \bibinfo{author}{\bibfnamefont{C.}~\bibnamefont{{Blake}}},
  \bibinfo{author}{\bibfnamefont{M.}~\bibnamefont{{Colless}}},
  \bibinfo{author}{\bibfnamefont{D.~H.} \bibnamefont{{Jones}}},
  \bibnamefont{et~al.}, \bibinfo{journal}{\mnras}
  \textbf{\bibinfo{volume}{416}}, \bibinfo{pages}{3017} (\bibinfo{year}{2011}),
  \eprint{1106.3366}.

\bibitem[{\citenamefont{{Blake} et~al.}(2012)\citenamefont{{Blake}, {Brough},
  {Colless}, {Contreras}, {Couch}, {Croom}, {Croton}, {Davis}, {Drinkwater},
  {Forster} et~al.}}]{2012MNRAS.425..405B}
\bibinfo{author}{\bibfnamefont{C.}~\bibnamefont{{Blake}}},
  \bibinfo{author}{\bibfnamefont{S.}~\bibnamefont{{Brough}}},
  \bibinfo{author}{\bibfnamefont{M.}~\bibnamefont{{Colless}}},
  \bibinfo{author}{\bibfnamefont{C.}~\bibnamefont{{Contreras}}},
  \bibnamefont{et~al.}, \bibinfo{journal}{\mnras}
  \textbf{\bibinfo{volume}{425}}, \bibinfo{pages}{405} (\bibinfo{year}{2012}),
  \eprint{1204.3674}.

\bibitem[{\citenamefont{{Padmanabhan} et~al.}(2012)\citenamefont{{Padmanabhan},
  {Xu}, {Eisenstein}, {Scalzo}, {Cuesta}, {Mehta}, and
  {Kazin}}}]{2012MNRAS.427.2132P}
\bibinfo{author}{\bibfnamefont{N.}~\bibnamefont{{Padmanabhan}}},
  \bibinfo{author}{\bibfnamefont{X.}~\bibnamefont{{Xu}}},
  \bibinfo{author}{\bibfnamefont{D.~J.} \bibnamefont{{Eisenstein}}},
  \bibinfo{author}{\bibfnamefont{R.}~\bibnamefont{{Scalzo}}},
  \bibnamefont{et~al.}, \bibinfo{journal}{\mnras}
  \textbf{\bibinfo{volume}{427}}, \bibinfo{pages}{2132} (\bibinfo{year}{2012}),
  \eprint{1202.0090}.

\bibitem[{\citenamefont{{Anderson} et~al.}(2012)\citenamefont{{Anderson},
  {Aubourg}, {Bailey}, {Bizyaev}, {Blanton}, {Bolton}, {Brinkmann},
  {Brownstein}, {Burden}, {Cuesta} et~al.}}]{2012MNRAS.427.3435A}
\bibinfo{author}{\bibfnamefont{L.}~\bibnamefont{{Anderson}}},
  \bibinfo{author}{\bibfnamefont{E.}~\bibnamefont{{Aubourg}}},
  \bibinfo{author}{\bibfnamefont{S.}~\bibnamefont{{Bailey}}},
  \bibinfo{author}{\bibfnamefont{D.}~\bibnamefont{{Bizyaev}}},
  \bibnamefont{et~al.}, \bibinfo{journal}{\mnras}
  \textbf{\bibinfo{volume}{427}}, \bibinfo{pages}{3435} (\bibinfo{year}{2012}),
  \eprint{1203.6594}.

\bibitem[{\citenamefont{{Planck Collaboration}
  et~al.}(2013{\natexlab{b}})\citenamefont{{Planck Collaboration}, {Ade},
  {Aghanim}, {Armitage-Caplan}, {Arnaud}, {Ashdown}, {Atrio-Barandela},
  {Aumont}, {Baccigalupi}, {Banday} et~al.}}]{2013arXiv1303.5076P}
\bibinfo{author}{\bibnamefont{{Planck Collaboration}}},
  \bibinfo{author}{\bibfnamefont{P.~A.~R.} \bibnamefont{{Ade}}},
  \bibinfo{author}{\bibfnamefont{N.}~\bibnamefont{{Aghanim}}},
  \bibinfo{author}{\bibfnamefont{C.}~\bibnamefont{{Armitage-Caplan}}},
  \bibnamefont{et~al.}, \bibinfo{journal}{ArXiv e-prints}
  (\bibinfo{year}{2013}{\natexlab{b}}), \eprint{1303.5076}.

\bibitem[{\citenamefont{{Planck Collaboration}
  et~al.}(2013{\natexlab{c}})\citenamefont{{Planck Collaboration}, {Ade},
  {Aghanim}, {Armitage-Caplan}, {Arnaud}, {Ashdown}, {Atrio-Barandela},
  {Aumont}, {Baccigalupi}, {Banday} et~al.}}]{2013arXiv1303.5080P}
\bibinfo{author}{\bibnamefont{{Planck Collaboration}}},
  \bibinfo{author}{\bibfnamefont{P.~A.~R.} \bibnamefont{{Ade}}},
  \bibinfo{author}{\bibfnamefont{N.}~\bibnamefont{{Aghanim}}},
  \bibinfo{author}{\bibfnamefont{C.}~\bibnamefont{{Armitage-Caplan}}},
  \bibnamefont{et~al.}, \bibinfo{journal}{ArXiv e-prints}
  (\bibinfo{year}{2013}{\natexlab{c}}), \eprint{1303.5080}.

\bibitem[{\citenamefont{{Benson} et~al.}(2013)\citenamefont{{Benson}, {de
  Haan}, {Dudley}, {Reichardt}, {Aird}, {Andersson}, {Armstrong}, {Ashby},
  {Bautz}, {Bayliss} et~al.}}]{2013ApJ...763..147B}
\bibinfo{author}{\bibfnamefont{B.~A.} \bibnamefont{{Benson}}},
  \bibinfo{author}{\bibfnamefont{T.}~\bibnamefont{{de Haan}}},
  \bibinfo{author}{\bibfnamefont{J.~P.} \bibnamefont{{Dudley}}},
  \bibinfo{author}{\bibfnamefont{C.~L.} \bibnamefont{{Reichardt}}},
  \bibnamefont{et~al.}, \bibinfo{journal}{\apj} \textbf{\bibinfo{volume}{763}},
  \bibinfo{eid}{147} (\bibinfo{year}{2013}), \eprint{1112.5435}.

\bibitem[{\citenamefont{{Hasselfield} et~al.}(2013)\citenamefont{{Hasselfield},
  {Hilton}, {Marriage}, {Addison}, {Barrientos}, {Battaglia}, {Battistelli},
  {Bond}, {Crichton}, {Das} et~al.}}]{2013JCAP...07..008H}
\bibinfo{author}{\bibfnamefont{M.}~\bibnamefont{{Hasselfield}}},
  \bibinfo{author}{\bibfnamefont{M.}~\bibnamefont{{Hilton}}},
  \bibinfo{author}{\bibfnamefont{T.~A.} \bibnamefont{{Marriage}}},
  \bibinfo{author}{\bibfnamefont{G.~E.} \bibnamefont{{Addison}}},
  \bibnamefont{et~al.}, \bibinfo{journal}{\jcap} \textbf{\bibinfo{volume}{7}},
  \bibinfo{eid}{008} (\bibinfo{year}{2013}), \eprint{1301.0816}.

\bibitem[{\citenamefont{{Vikhlinin} et~al.}(2009)\citenamefont{{Vikhlinin},
  {Kravtsov}, {Burenin}, {Ebeling}, {Forman}, {Hornstrup}, {Jones}, {Murray},
  {Nagai}, {Quintana} et~al.}}]{vik09}
\bibinfo{author}{\bibfnamefont{A.}~\bibnamefont{{Vikhlinin}}},
  \bibinfo{author}{\bibfnamefont{A.~V.} \bibnamefont{{Kravtsov}}},
  \bibinfo{author}{\bibfnamefont{R.~A.} \bibnamefont{{Burenin}}},
  \bibinfo{author}{\bibfnamefont{H.}~\bibnamefont{{Ebeling}}},
  \bibnamefont{et~al.}, \bibinfo{journal}{\apj} \textbf{\bibinfo{volume}{692}},
  \bibinfo{pages}{1060} (\bibinfo{year}{2009}), \eprint{0812.2720}.

\bibitem[{\citenamefont{{Rozo} et~al.}(2010)\citenamefont{{Rozo}, {Wechsler},
  {Rykoff}, {Annis}, {Becker}, {Evrard}, {Frieman}, {Hansen}, {Hao}, {Johnston}
  et~al.}}]{2010ApJ...708..645R}
\bibinfo{author}{\bibfnamefont{E.}~\bibnamefont{{Rozo}}},
  \bibinfo{author}{\bibfnamefont{R.~H.} \bibnamefont{{Wechsler}}},
  \bibinfo{author}{\bibfnamefont{E.~S.} \bibnamefont{{Rykoff}}},
  \bibinfo{author}{\bibfnamefont{J.~T.} \bibnamefont{{Annis}}},
  \bibnamefont{et~al.}, \bibinfo{journal}{\apj} \textbf{\bibinfo{volume}{708}},
  \bibinfo{pages}{645} (\bibinfo{year}{2010}), \eprint{0902.3702}.

\bibitem[{\citenamefont{{Planck Collaboration}
  et~al.}(2013{\natexlab{d}})\citenamefont{{Planck Collaboration}, {Ade},
  {Aghanim}, {Armitage-Caplan}, {Arnaud}, {Ashdown}, {Atrio-Barandela},
  {Aumont}, {Baccigalupi}, {Banday} et~al.}}]{2013arXiv1303.5077P}
\bibinfo{author}{\bibnamefont{{Planck Collaboration}}},
  \bibinfo{author}{\bibfnamefont{P.~A.~R.} \bibnamefont{{Ade}}},
  \bibinfo{author}{\bibfnamefont{N.}~\bibnamefont{{Aghanim}}},
  \bibinfo{author}{\bibfnamefont{C.}~\bibnamefont{{Armitage-Caplan}}},
  \bibnamefont{et~al.}, \bibinfo{journal}{ArXiv e-prints}
  (\bibinfo{year}{2013}{\natexlab{d}}), \eprint{1303.5077}.

\bibitem[{\citenamefont{van Engelen et~al.}(2012)\citenamefont{van Engelen,
  Keisler, Zahn, Aird, Benson et~al.}}]{vanEngelen:2012va}
\bibinfo{author}{\bibfnamefont{A.}~\bibnamefont{van Engelen}},
  \bibinfo{author}{\bibfnamefont{R.}~\bibnamefont{Keisler}},
  \bibinfo{author}{\bibfnamefont{O.}~\bibnamefont{Zahn}},
  \bibinfo{author}{\bibfnamefont{K.}~\bibnamefont{Aird}}, \bibnamefont{et~al.},
  \bibinfo{journal}{Astrophys.J.} \textbf{\bibinfo{volume}{756}},
  \bibinfo{pages}{142} (\bibinfo{year}{2012}), \eprint{1202.0546}.

\bibitem[{\citenamefont{{Kilbinger} et~al.}(2013)\citenamefont{{Kilbinger},
  {Fu}, {Heymans}, {Simpson}, {Benjamin}, {Erben}, {Harnois-D{\'e}raps},
  {Hoekstra}, {Hildebrandt}, {Kitching} et~al.}}]{2013MNRAS.430.2200K}
\bibinfo{author}{\bibfnamefont{M.}~\bibnamefont{{Kilbinger}}},
  \bibinfo{author}{\bibfnamefont{L.}~\bibnamefont{{Fu}}},
  \bibinfo{author}{\bibfnamefont{C.}~\bibnamefont{{Heymans}}},
  \bibinfo{author}{\bibfnamefont{F.}~\bibnamefont{{Simpson}}},
  \bibnamefont{et~al.}, \bibinfo{journal}{\mnras}
  \textbf{\bibinfo{volume}{430}}, \bibinfo{pages}{2200} (\bibinfo{year}{2013}),
  \eprint{1212.3338}.

\bibitem[{\citenamefont{Smith et~al.}(2003)}]{Smith:2002dz}
\bibinfo{author}{\bibfnamefont{R.}~\bibnamefont{Smith}} \bibnamefont{et~al.}
  (\bibinfo{collaboration}{Virgo Consortium}), \bibinfo{journal}{\mnras}
  \textbf{\bibinfo{volume}{341}}, \bibinfo{pages}{1311} (\bibinfo{year}{2003}),
  \eprint{astro-ph/0207664}.

\bibitem[{\citenamefont{Takahashi et~al.}(2012)\citenamefont{Takahashi, Sato,
  Nishimichi, Taruya, and Oguri}}]{Takahashi:2012em}
\bibinfo{author}{\bibfnamefont{R.}~\bibnamefont{Takahashi}},
  \bibinfo{author}{\bibfnamefont{M.}~\bibnamefont{Sato}},
  \bibinfo{author}{\bibfnamefont{T.}~\bibnamefont{Nishimichi}},
  \bibinfo{author}{\bibfnamefont{A.}~\bibnamefont{Taruya}},
  \bibnamefont{et~al.}, \bibinfo{journal}{Astrophys.J.}
  \textbf{\bibinfo{volume}{761}}, \bibinfo{pages}{152} (\bibinfo{year}{2012}),
  \eprint{1208.2701}.

\bibitem[{\citenamefont{{Planck collaboration}
  et~al.}(2013)\citenamefont{{Planck collaboration}, {Ade}, {Aghanim},
  {Armitage-Caplan}, {Arnaud}, {Ashdown}, {Atrio-Barandela}, {Aumont},
  {Baccigalupi}, {Banday} et~al.}}]{2013arXiv1303.5075P}
\bibinfo{author}{\bibnamefont{{Planck collaboration}}},
  \bibinfo{author}{\bibfnamefont{P.~A.~R.} \bibnamefont{{Ade}}},
  \bibinfo{author}{\bibfnamefont{N.}~\bibnamefont{{Aghanim}}},
  \bibinfo{author}{\bibfnamefont{C.}~\bibnamefont{{Armitage-Caplan}}},
  \bibnamefont{et~al.}, \bibinfo{journal}{ArXiv e-prints}
  (\bibinfo{year}{2013}), \eprint{1303.5075}.

\bibitem[{\citenamefont{{Wyman} et~al.}(2013)\citenamefont{{Wyman}, {Rudd},
  {Vanderveld}, and {Hu}}}]{2013arXiv1307.7715W}
\bibinfo{author}{\bibfnamefont{M.}~\bibnamefont{{Wyman}}},
  \bibinfo{author}{\bibfnamefont{D.~H.} \bibnamefont{{Rudd}}},
  \bibinfo{author}{\bibfnamefont{R.~A.} \bibnamefont{{Vanderveld}}},
  \bibnamefont{and} \bibinfo{author}{\bibfnamefont{W.}~\bibnamefont{{Hu}}},
  \bibinfo{journal}{ArXiv e-prints}  (\bibinfo{year}{2013}),
  \eprint{1307.7715}.

\bibitem[{\citenamefont{{Riess} et~al.}(2011)\citenamefont{{Riess}, {Macri},
  {Casertano}, {Lampeitl}, {Ferguson}, {Filippenko}, {Jha}, {Li}, and
  {Chornock}}}]{2011ApJ...730..119R}
\bibinfo{author}{\bibfnamefont{A.~G.} \bibnamefont{{Riess}}},
  \bibinfo{author}{\bibfnamefont{L.}~\bibnamefont{{Macri}}},
  \bibinfo{author}{\bibfnamefont{S.}~\bibnamefont{{Casertano}}},
  \bibinfo{author}{\bibfnamefont{H.}~\bibnamefont{{Lampeitl}}},
  \bibnamefont{et~al.}, \bibinfo{journal}{\apj} \textbf{\bibinfo{volume}{730}},
  \bibinfo{eid}{119} (\bibinfo{year}{2011}), \eprint{1103.2976}.

\bibitem[{\citenamefont{{Thomas} et~al.}(2010)\citenamefont{{Thomas},
  {Abdalla}, and {Lahav}}}]{2010PhRvL.105c1301T}
\bibinfo{author}{\bibfnamefont{S.~A.} \bibnamefont{{Thomas}}},
  \bibinfo{author}{\bibfnamefont{F.~B.} \bibnamefont{{Abdalla}}},
  \bibnamefont{and} \bibinfo{author}{\bibfnamefont{O.}~\bibnamefont{{Lahav}}},
  \bibinfo{journal}{Phys. Rev. Lett.} \textbf{\bibinfo{volume}{105}},
  \bibinfo{eid}{031301} (\bibinfo{year}{2010}), \eprint{0911.5291}.

\bibitem[{\citenamefont{{Riemer-S{\o}rensen}
  et~al.}(2013)\citenamefont{{Riemer-S{\o}rensen}, {Parkinson}, and
  {Davis}}}]{2013arXiv1306.4153R}
\bibinfo{author}{\bibfnamefont{S.}~\bibnamefont{{Riemer-S{\o}rensen}}},
  \bibinfo{author}{\bibfnamefont{D.}~\bibnamefont{{Parkinson}}},
  \bibnamefont{and} \bibinfo{author}{\bibfnamefont{T.~M.}
  \bibnamefont{{Davis}}}, \bibinfo{journal}{ArXiv e-prints}
  (\bibinfo{year}{2013}), \eprint{1306.4153}.

\bibitem[{\citenamefont{{S{\'a}nchez} and {Cole}}(2008)}]{2008MNRAS.385..830S}
\bibinfo{author}{\bibfnamefont{A.~G.} \bibnamefont{{S{\'a}nchez}}}
  \bibnamefont{and} \bibinfo{author}{\bibfnamefont{S.}~\bibnamefont{{Cole}}},
  \bibinfo{journal}{\mnras} \textbf{\bibinfo{volume}{385}},
  \bibinfo{pages}{830} (\bibinfo{year}{2008}), \eprint{0708.1517}.

\bibitem[{\citenamefont{{Hamann} and {Hasenkamp}}(2013)}]{2013arXiv1308.3255H}
\bibinfo{author}{\bibfnamefont{J.}~\bibnamefont{{Hamann}}} \bibnamefont{and}
  \bibinfo{author}{\bibfnamefont{J.}~\bibnamefont{{Hasenkamp}}},
  \bibinfo{journal}{ArXiv e-prints}  (\bibinfo{year}{2013}),
  \eprint{1308.3255}.

\bibitem[{\citenamefont{{Lewis} et~al.}(2000)\citenamefont{{Lewis},
  {Challinor}, and {Lasenby}}}]{2000ApJ...538..473L}
\bibinfo{author}{\bibfnamefont{A.}~\bibnamefont{{Lewis}}},
  \bibinfo{author}{\bibfnamefont{A.}~\bibnamefont{{Challinor}}},
  \bibnamefont{and}
  \bibinfo{author}{\bibfnamefont{A.}~\bibnamefont{{Lasenby}}},
  \bibinfo{journal}{\apj} \textbf{\bibinfo{volume}{538}}, \bibinfo{pages}{473}
  (\bibinfo{year}{2000}), \eprint{arXiv:astro-ph/9911177}.

\bibitem[{\citenamefont{{Lewis} and {Bridle}}(2002)}]{2002PhRvD..66j3511L}
\bibinfo{author}{\bibfnamefont{A.}~\bibnamefont{{Lewis}}} \bibnamefont{and}
  \bibinfo{author}{\bibfnamefont{S.}~\bibnamefont{{Bridle}}},
  \bibinfo{journal}{\prd} \textbf{\bibinfo{volume}{66}}, \bibinfo{eid}{103511}
  (\bibinfo{year}{2002}), \eprint{arXiv:astro-ph/0205436}.

\end{thebibliography}
\end{document}